\documentclass[twoside,twocolumn,9pt]{article}
\usepackage{extsizes}
\usepackage[super,sort&compress,comma]{natbib} 
\usepackage[version=3]{mhchem}
\usepackage[left=1.5cm, right=1.5cm, top=1.785cm, bottom=2.0cm]{geometry}
\usepackage{balance}
\usepackage{mathptmx}
\usepackage{sectsty}
\usepackage{graphicx} 
\usepackage{lastpage}
\usepackage[format=plain,justification=justified,singlelinecheck=false,font={stretch=1.125,small,sf},labelfont=bf,labelsep=space]{caption}
\usepackage{float}
\usepackage{fancyhdr}
\usepackage{fnpos}
\usepackage[english]{babel}
\addto{\captionsenglish}{%
  
}
\usepackage{array}
\usepackage{droidsans}
\usepackage{charter}
\usepackage[T1]{fontenc}
\usepackage[usenames,dvipsnames]{xcolor}
\usepackage{setspace}
\usepackage[compact]{titlesec}
\usepackage{xr}
\usepackage{hyperref}
\usepackage{cuted}
\usepackage{flushend}
\usepackage{lipsum}
\usepackage{dblfloatfix}    
\usepackage{subfiles} 
\usepackage{epstopdf}
\definecolor{cream}{RGB}{222,217,201}
\begin{document}
\pagestyle{fancy}
\thispagestyle{plain}
\fancypagestyle{plain}{
\renewcommand{\headrulewidth}{0pt}
}
\makeFNbottom
\makeatletter
\renewcommand\LARGE{\@setfontsize\LARGE{15pt}{17}}
\renewcommand\Large{\@setfontsize\Large{12pt}{14}}
\renewcommand\large{\@setfontsize\large{10pt}{12}}
\renewcommand\footnotesize{\@setfontsize\footnotesize{7pt}{10}}
\renewcommand\scriptsize{\@setfontsize\scriptsize{7pt}{7}}
\makeatother
\renewcommand{\thefootnote}{\fnsymbol{footnote}}
\renewcommand\footnoterule{\vspace*{1pt}%
\color{cream}\hrule width 3.5in height 0.4pt \color{black} \vspace*{5pt}} 
\setcounter{secnumdepth}{5}
\makeatletter 
\renewcommand\@biblabel[1]{#1}            
\renewcommand\@makefntext[1]%
{\noindent\makebox[0pt][r]{\@thefnmark\,}#1}
\makeatother 
\renewcommand{\figurename}{\small{Fig.}~}
\sectionfont{\sffamily\Large}
\subsectionfont{\normalsize}
\subsubsectionfont{\bf}
\setstretch{1.125} 
\setlength{\skip\footins}{0.8cm}
\setlength{\footnotesep}{0.25cm}
\setlength{\jot}{10pt}
\titlespacing*{\section}{0pt}{4pt}{4pt}
\titlespacing*{\subsection}{0pt}{15pt}{1pt}
\fancyhead{}
\renewcommand{\headrulewidth}{0pt} 
\renewcommand{\footrulewidth}{0pt}
\setlength{\arrayrulewidth}{1pt}
\setlength{\columnsep}{6.5mm}
\setlength\bibsep{1pt}
\makeatletter 
\newlength{\figrulesep} 
\setlength{\figrulesep}{0.5\textfloatsep} 
\newcommand{\topfigrule}{\vspace*{-1pt}%
\noindent{\color{cream}\rule[-\figrulesep]{\columnwidth}{1.5pt}} }
\newcommand{\botfigrule}{\vspace*{-2pt}%
\noindent{\color{cream}\rule[\figrulesep]{\columnwidth}{1.5pt}} }
\newcommand{\dblfigrule}{\vspace*{-1pt}%
\noindent{\color{cream}\rule[-\figrulesep]{\textwidth}{1.5pt}} }
\makeatother
\twocolumn[
\begin{@twocolumnfalse}
\vspace{1em}
\sffamily

\noindent\LARGE{\textbf{Percolation in Networks of Liquid Diodes}} \\
  \vspace{0.3cm} \\
  \noindent\large{Camilla Sammartino$^{a}$, Yair Shokef$^{a,b,c,d}$, and Bat-El Pinchasik$^{a,b}$} \\
\end{@twocolumnfalse} \vspace{0.6cm}
]
\renewcommand*\rmdefault{bch}\normalfont\upshape
\rmfamily
\section*{}
\vspace{-1cm}

\footnotetext{\textit{$^{a}$~School of Mechanical Engineering, Tel Aviv University, Tel Aviv 69978, Israel. E-mail: camillas@mail.tau.ac.il , shokef@tau.ac.il , pinchasik@tauex.tau.ac.il}}

\footnotetext{\textit{$^{b}$~Center for Physics and Chemistry of Living Systems, Tel Aviv University, Tel Aviv 69978, Israel.}}

\footnotetext{\textit{$^{c}$~Center for Computational Molecular and Materials Science, and Center for Physics and Chemistry of Living Systems, Tel Aviv University, Tel Aviv 69978, Israel.}}

\footnotetext{\textit{$^{d}$~International Institute for Sustainability with Knotted Chiral Meta Matter, Hiroshima University, Japan.}}



\sffamily{\textbf
{Liquid diodes are surface structures that facilitate the flow of liquids in a specific direction. When these structures are within the capillary regime, they promote liquid transport without the need for external forces. In nature, they are used to increase water collection and uptake, reproduction, and feeding. While nature offers various one-dimensional channels for unidirectional transport, networks with directional properties are exceptional and typically limited to millimeters or a few centimeters. In this study, we simulate, design and 3D print liquid diode networks consisting of hundreds of unit cells. We provide structural and wettability guidelines for directional transport of liquids through these networks, and introduce percolation theory in order to identify the threshold between a connected network, which allows fluid to reach specific points, and a disconnected network. By constructing well-defined networks that combine uni- and bi-directional pathways, we experimentally demonstrate the applicability of models describing isotropically directed percolation. By varying the surface structure and the solid-liquid interfacial tension, we precisely control the portion of liquid diodes and bidirectional connections in the network and follow the flow evolution. We are, therefore, able to accurately predict the network permeability and the liquid's final state. These guidelines are highly promising for the development of structures for spontaneous, yet predictable, directional liquid transport.}}


\rmfamily 



\section*{Introduction}
\label{intro}

Percolation theory is used to describe critical phenomena in multiple types of complex physical systems\cite{stauffer_introduction_2018} such as flow through porous or granular media\cite{Complex_percolation_theory, sahimi_applications_2023}, electrical conductivity\cite{electrical_graphene_2018}, spreading of fires\cite{abades_fire_2014}, vascular networks\cite{chimal-eguia_vascular_2020}, biomolecular transport\cite{gomez_mechanical_2019, gomez_target_2020}, jamming of particulate systems
\cite{TBF_2006, ghosh_jamming_2014, segall_jamming_2016}, and even the formation and release of traffic jams\cite{zeng_switch_2019}. These phenomena can be described by different percolation models\cite{stauffer_introduction_2018, li_percolation_2021}, through the formation of connected clusters and networks. Well-defined, predictable experimental realizations of these models, however, still remain an open field of research, with potential applications in fluidics, electronics, power grids, epidemics, and biology\cite{dorogovtsev_critical_2008}. 

To date, experimental efforts have focused on percolation of electrical conductivity\cite{lu_multiscale_2017, conductivity_2019, haghgoo_novel_2022, mazaheri_modeling_2022, zhang_multi-functional_2020}, mainly using the bond percolation model. In this model, the network is characterized by the probability $p$ of having a bond between two neighboring sites. Similarly, $p_0 = 1 - p$ is the probability of having a missing bond, namely a vacancy. If enough bonds are present, an infinite connected cluster is formed, the network is connected and percolates. The critical probability of existing bonds, $p_c$, defines the threshold between the non-percolating and percolating phases\cite{ziff_exact_2006}. 

In classic, or random, bond percolation, all bonds are bi-directional. Consequently, percolation through the network is isotropic. In directed percolation, on the other hand, the bonds are directional, and allow transport only in a preferred direction in the network\cite{obukhov_problem_1980, Hinreschsen_DP_review}. Thus, such a system is connected anisotropically. An example of such directional transport in nature is seen in the Texas horned lizard. This desert lizard increases its water collection and intake by passive water transport through a network of directional channels on its scales, directing the water to the mouth over distances of a few centimeters\cite{martin-palma_texas_2016, comanns_directional_2015}. 
Another, more recently introduced, percolation model corresponds to isotropically directed percolation\cite{redner_directed_1982, inui_1999, janssen_random_2000, zhou_crossover_2012, de_noronha_percolation_2018}. There, bonds in opposite directions exist with a total probability $p_1$, together with bi-directional bonds, with probability $p_2$, and vacancies with probability $p_0 = 1 - p_1 - p_2$. This type of percolation is governed by the probability to find a bond connecting two neighboring sites\cite{de_noronha_percolation_2018}, defined as $p_{nn}=\frac{p_1}{2}+p_2$. Isotropically directed percolation has been used to analyze traffic patterns in New York and London\cite{cogoni_stability_2021, verbavatz_one-way_2021}, but more experimental work is needed to further understand the applicability of the theory to real-life situations. 

\begin{figure*}[!b]
\centering
\includegraphics[scale=0.4]{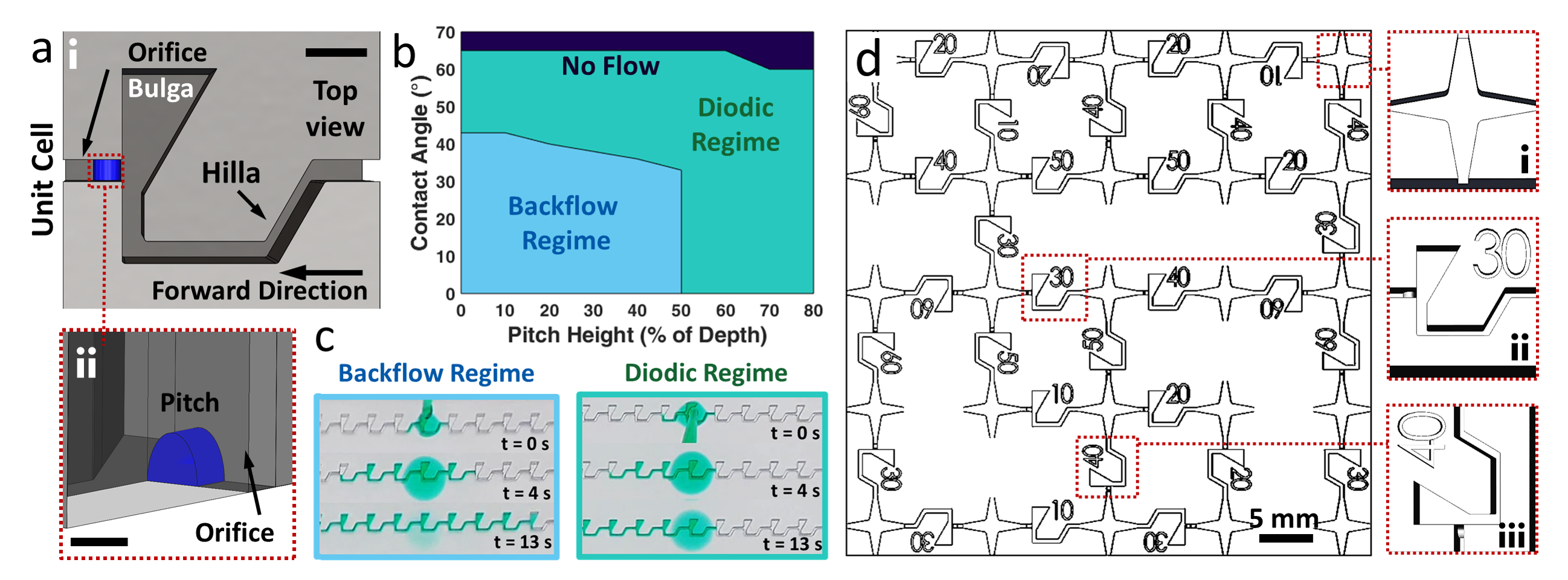}
\caption{Liquid diodes design and functioning. (a) Unit cell and main components (i). (ii) The pitch, an ellipsoid-shaped bump (blue). (b) Flow phase diagram, depending on the pitch height and liquid contact angle. Three main behaviors are observed: no flow (dark blue), diodic regime (in green) and backflow regime (in light blue). (c) Channels made of ten consecutive liquid diodes with 0\% (left) 40\% (right) pitch height, showing backflow and diodic regime, respectively. (d) Schematics of a 5$\times$5 isotropically directed network. The insets show the design of (i) a node, (ii) a 30\% pitch height bond oriented so that the forward direction is right to left, and (iii) a 40\% pitch height bond oriented so that the forward direction is top to bottom.}
\label{fig1}
\end{figure*}

In this study, we introduce an experimental realization of isotropically directed percolation using a network of three-dimensional (3D) printed network of liquid diodes. The liquid diodes comprise 3D surface structures that promote spontaneous uni-directional liquid flow in the capillary regime\cite{liquid_diodes}. We design a 2D network made of millimeter-size open channels\cite{sammartino_three-dimensional_2022, liquid_diodes} that set the bonds between neighboring sites. By tuning geometric features of the liquid diodes and the contact angle (CA) that the flowing liquid creates with the surface, we control the diodicity of the bonds. Namely, whether the flow through a bond is uni- or bi-directional. As a result, we are able to control the number of directional bonds, scan over different values of $p_{nn}$, switch between different percolation states and find the percolation threshold for each configuration of the network. Establishing a well-defined physical system that enables fine-tuning of percolation parameters enables us to gain new insights into the fundamentals of directional transport phenomena in general\cite{Feldmann_directional_2021}, and isotropically directed percolation specifically. This includes direct measurements of the flow dynamics through the network, and studying the influence of the system size and initial conditions of the fluid on its final state and configuration. In addition, controlling the percolation threshold and being able to tune the permeability of a network opens new horizons for designing microfluidic complex networks for mixing and separation\cite{feng_araucaria_2021, zhang_multi-functional_2020}, heat transfer\cite{gao_fluid_2022} and actuation\cite{droplet_actuation}.

\section*{Results and Discussion}

\subsection*{Liquid diodes networks: design and arrangement}

Figure~\ref{fig1} shows the liquid diodes used in this work\cite{sammartino_three-dimensional_2022} and how they are implemented to form 2D disordered isotropically directed networks. Each diode features an asymmetric geometry comprising four main components (Figure~\ref{fig1}a): the entrance channel (hilla), a central area (bulga), an exit channel (orifice), and, in blue, an ellipsoid-shaped bump (pitch), shown in more detail in the inset of Figure~\ref{fig1}a. The bulga is aligned perpendicularly to the orifice, creating a 90$^\circ$ expansion in the channel’s width. This expansion, together with the pitch, creates a pressure barrier that pins the liquid in the backward direction (left to right). The height of the pitch is defined in terms of percentage of the channel’s depth, and in our networks, it varies between 10\% and 60\%, in steps of 10\%. 

Different flow regimes arise, depending on the combination of the pitch height and the CA of the flowing liquid, summarized in the phase diagram shown in Figure~\ref{fig1}b. In green is the diodic regime, where the flow is unidirectional. Figure~\ref{fig1}c depicts timeframes of liquid flow in a channel made of several consecutive unit cells with 40\% pitch height and a CA of 45$^\circ$. By lowering the CA (i.e. increasing the liquid wettability\cite{Pinchasik_polymer_2014}), the diodes start exhibiting flow in the backward direction and the diodicity breaks (Figure~\ref{fig1}c, left). The dependence of the diodes' performance on the pitch height is associated with the local reduction and following expansion of the channel’s depth, caused by the pitch, when the liquid flows in the backward direction\cite{sammartino_three-dimensional_2022}. For example, with a pitch height of 40\%, the channel depth is locally reduced to 60\% its original value. After the pitch, the channel's depth returns to its full value (100\%). This creates an additional pressure barrier when liquid propagates in the backward direction and has to overcome the pitch (see Figure~\ref{figS1}). In the topmost part of the phase diagram, in dark blue, we observe no flow in either direction due to the poor wettability of the liquid.

\begin{figure*}[h]
\centering    
\includegraphics[scale=0.45]{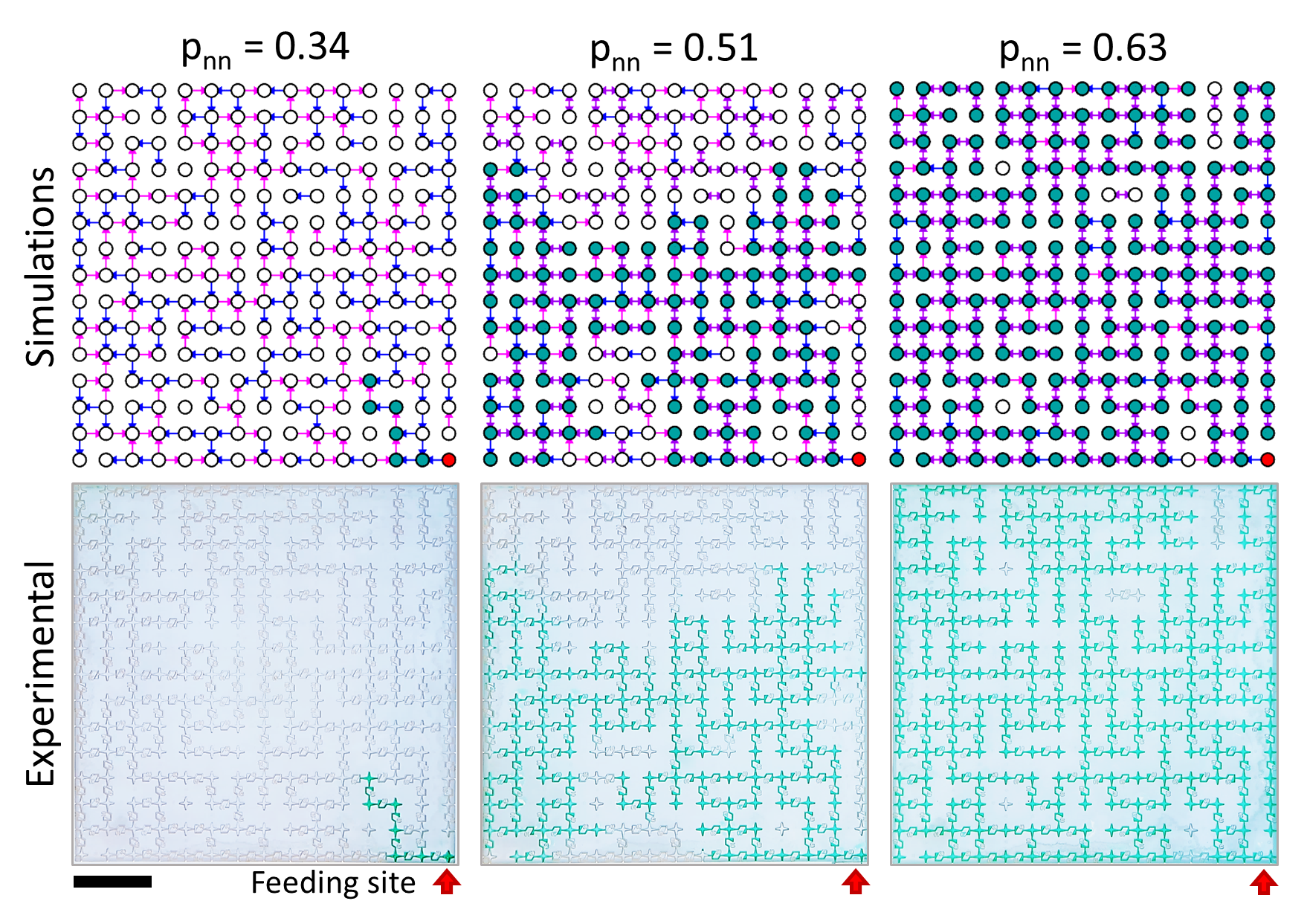}
\caption{Simulations (top) and experiments (bottom) of final percolation states for increasing values of $p_{nn}$, for a sample with $p_0$ = 0.31, starting from the same feeding site, highlighted in red. The scale bar corresponds to 3 mm. The percolation transition is clearly visible, and experiments show excellent agreement with simulations.}
\label{fig2}
\end{figure*}

Figure~\ref{fig1}d shows an isotropically directed 5$\times$5 network comprising diodes of different pitch heights and orientations. We created three networks, 15$\times$15 in size, with $p_0$ values of 0.2, 0.31 and 0.37, featuring, 336, 288 and 264 bonds, respectively, using diodes with pitch heights ranging from 10\% to 60\%, in steps of 10\%. Within a given sample, we have the same number of diodes per pitch height and orientation. Hence, at each site, the probability of a unidirectional bond in each direction is uniform and the network is isotropic. The distribution of vacancies and diodes, in each direction, is randomized (see Materials and Methods). As we change the CA of the flowing liquid, a fraction of the bonds switches behavior, according to the phase diagram in Figure~\ref{fig1}b, allowing us to scan over different values of $p_{nn}$ for each sample of a given $p_0$. This is illustrated in Figure~\ref{figS2}, using schematics of the three samples with different $p_0$, for increasing $p_{nn}$ values, or decreasing CAs. As $p_{nn}$ increases, the number of bidirectional bonds, indicated by magenta two-way arrows (Figure~\ref{figS1}), increases, creating a larger connected cluster.

\subsection*{Spontaneous directional flow in liquid diodes networks: simulations and experiments}

We first use numerical simulations to predict the flow pattern and the liquid permeability through the network. For each network configuration, we insert the "liquid" at a specific feeding site and let it propagate spontaneously, until the "liquid" fronts halt. Once the "liquid" reaches the final state, we extract the fraction of occupied sites $f$.
This was repeated for eight randomly chosen feeding sites, two for each edge of the network, and for each CA. The simulations included all the possible feeding sites along the edges of the network. We then verify our predictions by flow experiments in the 3D-printed liquid diodes networks.

Figure~\ref{fig2} shows simulations (upper row) and experiments (bottom row) of the liquid final states with increasing $p_{nn}$ (0.34, 0.51, 0.63), for $p_{0}$= 0.31. The red arrows denote the feeding sites (identical in all cases). When $p_{nn}$ increases, a larger portion of the network is covered with liquid, a result of the increasing number of bidirectional bonds. Videos S1, S2 and S3 show the propagation dynamics of the three experiments. This leads to a phase transition between a non-percolating (LHS) and a percolating state (RHS). The liquid distribution is isotropic. Namely, no preferred direction is observed, and the liquid spreads homogeneously in all directions. We find an excellent agreement between the simulations and experiments. Yet, a few local failures in the diodes were observed due to pressure buildup. This additional pressure renders unidirectional bonds bidirectional, promoting flow in the backward direction and local breakdown of diodicity\cite{davletshin_bidirectional_2022}. 

\begin{figure}[h!]
\centering\includegraphics[scale=0.44]{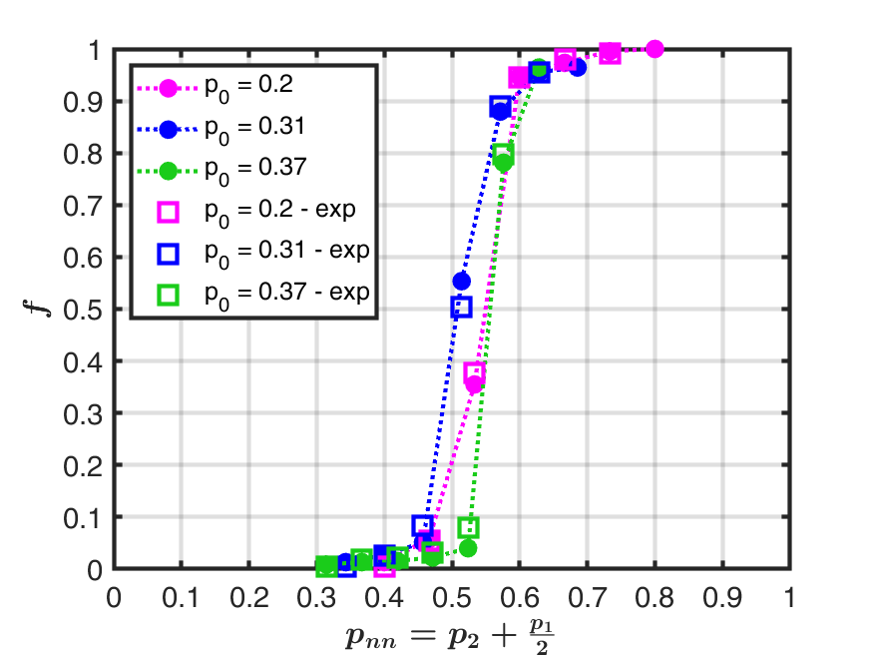}
\caption{Fraction $f$ of occupied sites as a function of $p_{nn}$, for 15$\times$15 networks with increasing values of $p_0$ = 0.2, 0.31 and 0.37. Simulations (circles) and experiments (diamonds) show excellent agreement. Error bars in the experimental curves are smaller than the markers. The data collapse presents some scatter due to the limited system size.}
\label{fig3}
\end{figure}

\begin{figure*}[t!]
\centering
\includegraphics[scale=0.5]{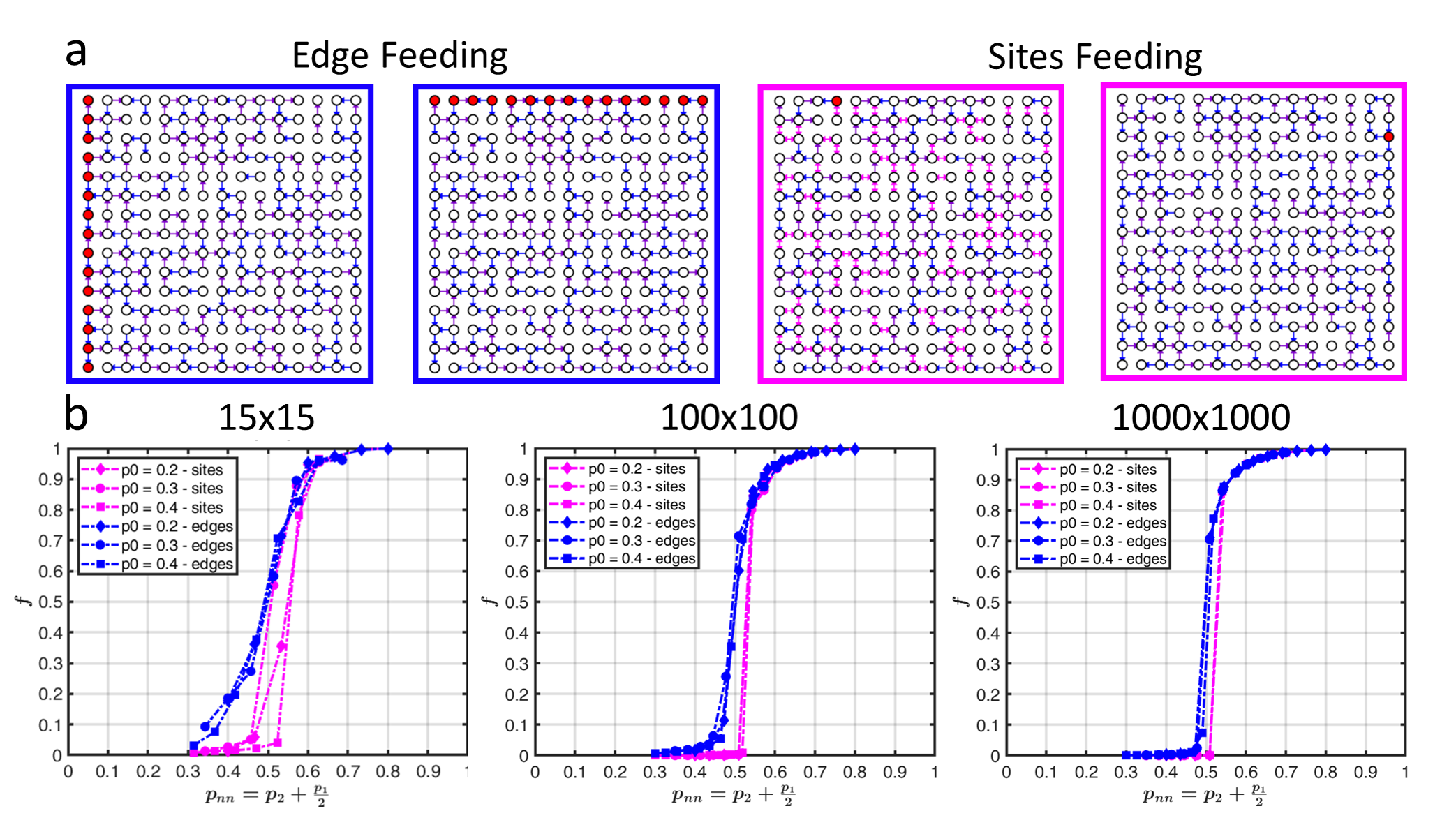}
\caption{(a) Schematic representation of edge feeding (left) and sites feeding (right). (b) Simulations showing the effect of system size and feeding method on percolation curves, for a 15$\times$15, 100$\times$100 and 1000$\times$1000 isotropically directed network. As the system size increases, the phase transition becomes sharper and the discrepancy between the feeding methods negligible.}
\label{fig4}
\end{figure*}

We now examine the dependence of $f$, the fraction of occupied sites, on $p_{nn}$, as shown in Figure~\ref{fig3}. We observe the typical S-shape of percolation curves\cite{stauffer_introduction_2018}. The three curves, for the three different $p_0$ values, show identical behavior and a notable degree of collapse. Simulations (circles) and experiments (squares) show excellent agreement. For each set of experiments, with a specific CA (i.e., $p_{nn}$), the fraction of occupied sites is averaged by taking the median of all experiments. This was done to  account for discrepancies in the outcomes of experiments with the same $p_{nn}$ but different feeding sites. In fact, not all sites on the border result necessarily in a similar connected path and, hence, a similar $f$. Using the median reflects the distribution of the results and gives each experiment the appropriate weight, according to how often a specific outcome occurred. 

The percolation threshold lies around $p_{nn}=\frac{1}{2}$, as expected for the square lattice\cite{kesten_squarelattice_1980, de_noronha_percolation_2018}. The variability seen in the data collapse is due to the moderate system size of the 3D-printed networks\cite{coniglio_cluster_1982, li_finite-size_2009}. The numerical simulations show that increasing the system size results in a better data collapse and a sharper percolation transition. This is manifested in Figure~\ref{fig4} by the magenta plots, for system sizes of 15$\times$15, 100$\times$100 and 1000$\times$1000. Each different marker shape represents a different $p_{0}$ value. 

Finally, we investigate the effect of different feeding protocols on the shape of the percolation curves for finite system sizes. The blue plots in Figure~\ref{fig4} represent percolation curves for increasing system size, obtained through numerical simulations, in which an entire edge of the network was fed rather than a single site on the edge. In this way, only four different initial conditions are averaged, in contrast to fifty-six in the case of single feeding sites. This feeding method results in a less sharp transition and a more symmetrical S-shape of the curve. The discrepancy between the two methods, more acute for small $p_{nn}$ values, becomes less significant as the system size increases. Indeed, feeding the system through an entire edge results in an increased number of occupied sites for small $p_{nn}$ values. This increase is significant for moderate (15$\times$15) to medium (100$\times$100) system sizes but becomes minute for larger (1000$\times$1000) networks.

\section*{Conclusions}

In this work, we successfully conducted isotropically directed percolation experiments of fluid flow in 2D networks of liquid diodes. Fine-tuning of the liquid diodes geometry enabled us to control and manipulate the network's statistical properties. Namely, to precisely control the amount of unidirectional and bidirectional bonds, and therefore, the flow patterns and propagation through the network. The excellent agreement between numerical simulations and experiments for finite-size systems of 15$\times$15 sites validates the integrity of the simulations for larger scales. We have gained new insights into the impact of the feeding method (initial conditions) on the percolation transition. When feeding the system from a single site, percolation curves are steeper and the phase transition is more abrupt than feeding the system from an entire edge. This becomes less significant as we increase the system size and should be taken into account when analyzing or modeling real-life systems of moderate sizes. Additionally, we showed controlled passive transport of liquids in two dimensions over distances of about 15~cm. This is a testimony to the potential of liquid diodes not only as tools to investigate fundamental scientific questions, but also to fabricate devices with a high degree of design flexibility. Further work, using different fabrication methods, may allow us to scale down the dimensions of the unit cells and sites, resulting in bigger networks, featuring thousands of bonds.      

\section*{Materials and Methods}

\subsection*{3D Printed Sample Design}

The liquid diodes design is based on our previous work \cite{sammartino_three-dimensional_2022}. The percentage of the pitch height was marked next to each unit cell for better visualization of the final sample. To design 2D networks, a computerized script was used to create a randomized isotropic distribution of diodes with various pitch heights for each value of $p_0$ (number of vacancies). Namely, for every pitch height, the number of diodes in each orientation (right, left, up, down) is identical. This way, from any random site within the network, there is a uniform probability of propagating in any direction. 

The nodes of the network are designed as a 4-point star with filleted edges (Figure 1d, i). This way, each junction to the neighboring diode, when present, is a wedge, and liquid is free to flow along the edges and around the corners. This proved to be more effective than a simple 90$^\circ$ cross design, as the liquid would get potentially pinned at the intersections. The size of each sample is 15X15 sites. This was the most convenient size in terms of fabrication and experiments. 

\subsection*{Sample and Material Preparation}

Samples were 3D-printed, using the ProJet® MJP 2500 Series by 3D Systems (Rock Hill, South Carolina, USA), a multi-jet 3D printer. The material used for the printing is the VisiJet® M2R-CL, a transparent polymer. Each sample, for each $p_0$ value, was printed twice to check repeatability. After printing, the parts were cleaned to remove the supporting wax material. First, the support wax was dissolved in canola oil at 60$^\circ$C. The sample was then placed in a second heated oil bath for finer wax removal. The oil residues were later washed away in soapy water at 60$^\circ$C (all-purpose liquid detergent soap, soap to water volume ratio corresponds to 0.3:1). A small brush was used to gently clean the residues in narrow voids, without damaging small features. Soap residues were thoroughly rinsed in deionized (DI) water, and the sample was dried in open air and lastly rinsed with Ethanol and dried with an air gun. In all experiments, we used dyed DI water obtained by adding green food coloring (Maimon’s, Be’er Sheva, Israel) in a volume ratio of 0.05:1 (dye to water). Small quantities of all-purpose liquid detergent soap, ranging from a volume ratio of 0.02:1 to 0.07:1, were added to the water in order to decrease the native contact angle of the liquid. Six solutions with different contact angles were prepared and used in the experiments.

The cleanliness of the sample surface was crucial for good performance and minimal unexpected pinning of the liquid. After each experiment, samples were rinsed with DI water and ethanol to remove all soap residues, and were then let to dry completely. 

\subsection*{Contact Angle Measurements}

The CA of the liquids was measured using a contact angle goniometer OCA25 by DataPhysics Instruments GmbH (Filderstadt, Germany). For each of the six liquids, five measurements were taken of 2 \textmu L drops from different areas of the surface. The CA was checked before each experiment. 

\subsection*{Percolation Experiments and Image Analysis}

For each sample and for each CA (thus $p_{nn}$ value), eight random sites along the outer edges, namely two sites per edge, were picked as the feeding sites for percolation experiments, for a total of 144 experiments. Liquid of a known contact angle was slowly poured into the feeding site using a Transferpette® S pipettes by BRAND GMBH + CO KG (Wertheim, Germany), in steps of 20 \textmu L. This method enabled us to avoid pressure buildup around the feeding site, with consequent possible failures of neighboring diodes. Experiments were recorded from above using a Panasonic DC-S1 camera with Sigma 70 mm F2.8 DG Macro lens. Screenshots of videos and the fraction of occupied sites were obtained using a MATLAB script, adjusted from the one used in our previous work\cite{sammartino_three-dimensional_2022}. The algorithm separates the final frame of the video into three RGB channels and subtracts the green from the red, highlighting only the liquid front with respect to the sample background. The frame is turned into a binary black-and-white image, which is divided into squares, each comprising a diode, a site, or a vacancy. For each square comprising a site, the mean of the central pixels is calculated. If the mean is larger than 0.3 (mostly white pixels), then the site is empty. If the mean is lower than 0.3, the site is counted as filled by the liquid. For each set of experiments for a specific $p_{nn}$ value, the median of the fraction of occupied sites over the eight experiments was taken. 

\subsection*{Numerical Simulations}

The numerical simulations reproduce the 3D-printed sample designs. For each CA, the distributions of the uni-directional and bi-directional bonds are well-defined, based on the phase diagram in Figure~\ref{fig1}b. The $p_{nn}$ value of each design is calculated and recorded. The network is then fed from a single site on the outer border and the occupancy (fraction of occupied bonds) is computed using a pass/no-pass function according to the directionality of each bond. This is repeated for every site on the border, and the median of all 56 experiments is calculated. The fraction $f$ of occupied sites is then plotted against the corresponding $p_{nn}$ value.

\section*{Author Contributions}

CS, YS and BP jointly conceived of the research. CS designed and prepared the samples, performed the experiments and simulations, and analyzed the results. YS provided theoretical guidance. BP provided experimental guidance. CS, YS and BP jointly wrote the paper.

\section*{Conflicts of interest}
 There are no conflicts to declare.

\section*{Acknowledgments}
This research was supported by the Israel Science Foundation (grant No. 1323/19).


\scriptsize{
\bibliography{Bibliography}
\bibliographystyle{rsc} }


\renewcommand{\thefigure}{S\arabic{figure}}
\setcounter{figure}{0}

\begin{figure*}[h]
\centering
\includegraphics[scale=0.5]{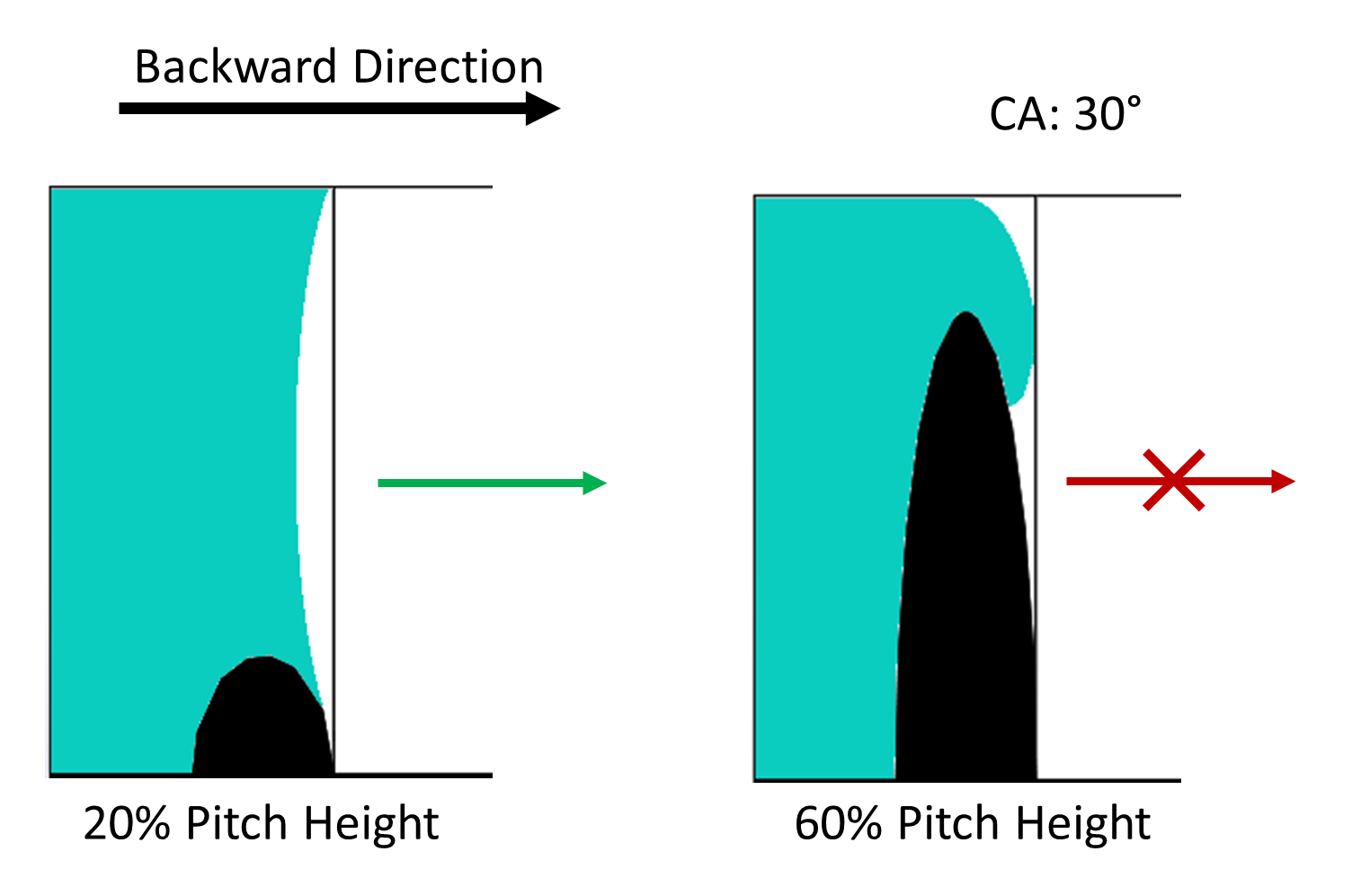}
\caption{Schematics with side view of the diodes' illustrating their functioning. A bigger pitch height results in a larger pressure barrier in the backward direction.}
\label{figS1}
\end{figure*}

\begin{figure*}[h]
\centering
\includegraphics[scale=0.4]{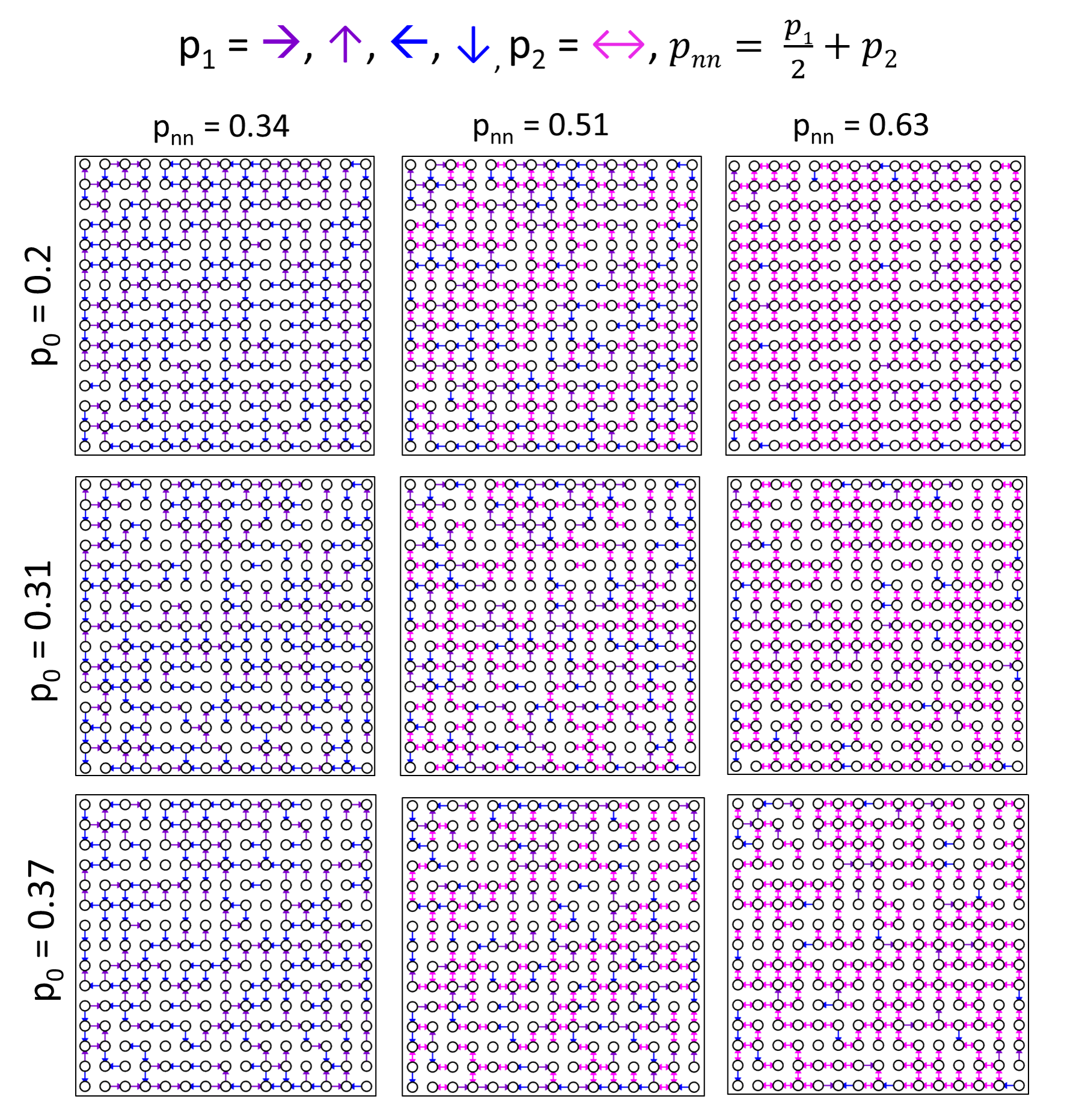}
\caption{Schematic representations of 15x15 networks with increasing $p_0$ values, for increasing values of $p_{nn}$. As $p_{nn}$ increases, a fraction of bonds becomes bidirectional (magenta double-pointed arrows), resulting in a bigger connected cluster.}
\label{figS2}
\end{figure*}

\end{document}